# Origin of the heavy elements in binary neutron-star mergers from a gravitational wave event

Daniel Kasen[1,2], Brian Metzger[3], Jennifer Barnes[3,4], Eliot Quataert[1] & Enrico Ramirez-Ruiz[5,6]

[1]Departments of Physics and Astronomy, and Theoretical Astrophysics Center, University of California, Berkeley, California 94720-7300, USA.

[2]Nuclear Science Division, Lawrence Berkeley National Laboratory, Berkeley, California 94720-8169, USA.

[3]Department of Physics and Columbia Astrophysics Laboratory, Columbia University, New York, New York 10027, USA.

[4]Einstein Fellow.

[5]Department of Astronomy, University of Santa Cruz, California, USA.

[6]DARK, Niels Bohr Institute, University of Copenhagen, Blegdamsvej 17, 2100 Copenhagen, Denmark.

**The cosmic origin of the elements heavier than iron has long been uncertain. Theoretical modelling[1–7] shows that the matter that is expelled in the violent merger of two neutron stars can assemble into heavy elements such as gold and platinum in a process known as rapid neutron capture (r-process) nucleosynthesis. The radioactive decay of isotopes of the heavy elements is predicted[8–12] to power a distinctive thermal glow (a 'kilonova'). The discovery of an electromagnetic counterpart to the gravitational-wave source[13] GW170817 represents the first opportunity to detect and scrutinize a sample of freshly synthesized r-process elements[14-18]. Here we report models that predict the detailed electromagnetic emission of kilonovae and enable the mass, velocity and composition of ejecta to be derived from the observations. We compare the models to the optical and infrared radiation associated with GW170817 event to argue that the observed source is a kilonova. We infer the presence of two distinct components of ejecta, one composed primarily of light (atomic mass number less than 140) and one of heavy (atomic mass number greater than 140) r-process elements. Inferring the ejected mass and a merger rate from GW170817 implies that such mergers are a dominant mode of r-process production in the Universe.**

The discovery[13] by the LIGO–Virgo experiments of gravitational waves from inspiralling neutron stars triggered an extensive campaign of follow-up observations, and the detection of counterpart emission across the electromagnetic spectrum. At optical and infrared wavelengths, the counterpart to GW170817 (originally announced by the Swope team[18] and called SSS17A, and hereafter referred to by its IAU designation AT 2017gfo) has properties that differ from previously known astrophysical transients. A day after the merger, the source was optically bright (with luminosity of about $10^8$ times that of the Sun at wavelengths of about 0.5 μm), but it faded rapidly within days[14-16]. Meanwhile, the emission in the infrared (1–3 μm) remained bright for nearly two weeks[15-19]. The spectra of AT 2017gfo were quasi-blackbody, suggestive of a thermal source[20,21].

The characteristics of AT 2017gfo—its rapid optical evolution[8], long-lived infrared emission[11,12] and luminosity consistent with radioactive r-process heating[9,10] resemble theoretical predictions for kilonovae. To explore this identification, we present here a survey of models of the radioactive aftermath of a neutron star merger. The key parameters of the models are the ejected mass $M$, characteristic expansion velocity $v_k$, and the composition of ejected matter. The ejecta is freely expanding (radius equal to the product of velocity and time, $R = vt$) and the density profile is described by a broken power law (see Methods). We synthesize model observables by numerically solving the Boltzmann equation for relativistic radiation transport in a radioactive plasma. We self-consistently calculate the thermal and ionization/excitation state of the ejecta and derive the wavelength-dependent opacity and emissivity using atomic-structure model data for multiple ions (see Methods). The validity of the transport method has been established by previous modelling of supernovae of all types.

We explore models motivated by general-relativistic simulations of mergers, which identify two distinct mechanisms for mass ejection (see Fig. 1). First, matter may be dynamically expelled on a timescale of milliseconds during the merger itself. Tidal forces peel matter from the surfaces of the approaching stars[4], flinging out tails of debris. As the stars come into contact, additional matter may be squeezed into the polar regions by shock heating at the interface[7,22]. The dynamical ejecta velocities are $v \approx 0.2c$–$0.3c$ (of the order of the escape velocity of a neutron star, where $c$ is the speed of light in vacuum) and the masses are in the range $M \approx 10^{-3} M_\odot – 10^{-2} M_\odot$, depending on the binary mass ratio and the uncertain equation of state. A second ejection mechanism can occur roughly a second after the merger, as matter hung up in an accretion disk around the central merged remnant is partially blown away in winds[5,6,23-26] with lower velocities of $0.05c–0.1c$ and masses of around $10^{-2} M_\odot – 10^{-1} M_\odot$.

As the ejecta decompresses from high densities, heavy elements are synthesized via the r-process[1–3,10]. If the matter is very neutron-rich (neutron mass fraction $\eta \geq 0.75$), repeated neutron captures forge the heaviest r-process elements ($58 \leq Z \leq 90$). If the ejecta is only moderately neutron-rich ($0.6 \leq \eta \leq 0.75$) nucleosynthesis stalls after producing only light r-process elements ($28 \leq Z \leq 58$). Tidal tail ejecta is expected to be predominantly cold and neutron-rich ($\eta \geq 0.8$) but the 'squeezed' polar ejecta and disk winds may have a lower $\eta$ distribution if they are subject to neutrino irradiation and weak interactions that convert neutrons into protons[5,7,27,28].

Hydrodynamic and nucleosynthetic processes abate several seconds after the merger, but the expanding cloud of ejecta is kept warm (temperatures of approximately $10^3–10^4$ K) for weeks by the decay of radioactive r-process isotopes, with an approximate heating rate[9] in units of ergs per second per gram:

$$\varepsilon_{\text{nuc}} \approx 10^{10}\, t^{-\alpha} \qquad (1)$$

where time $t$ is in days and $\alpha \approx 1.3$. The power-law time dependence is a statistical consequence of having numerous isotopes with half-lives approximately uniformly distributed in logarithmic time[8]. The overall level of heating, however, can vary by a factor of roughly 5, depending on the exact composition and uncertain nuclear data inputs.

Once the ejecta has expanded enough to becomes translucent, it releases thermal radiation. This occurs when the photon diffusion timescale $t_{\text{diff}} = \rho \kappa R^2/c$ becomes comparable to the elapsed time. Taking a constant opacity $\kappa$, constant expansion speed $v$, and uniform density $\rho = M/(4\pi R^3/3)$ gives a simple scaling for the duration of a kilonova light curve

$$t_{\text{lc}} \approx \left(\frac{3\kappa M}{4\pi cv}\right)^{\frac{1}{2}} \approx 2.7\,\text{days} \times \left(\frac{M}{0.01M}\right)^{\frac{1}{2}} \left(\frac{v}{0.1c}\right)^{-\frac{1}{2}} \left(\frac{\kappa}{1\,\text{cm}^2\,\text{g}^{-1}}\right)^{\frac{1}{2}} \quad (2)$$

The characteristic luminosity of the kilonova is approximately equal to the radioactive heating rate at this escape timescale

$$L_{\text{lc}} \approx M\varepsilon_{\text{nuc}}(t_{\text{lc}}) \approx 5 \times 10^{40}\,\text{erg}\,\text{s}^{-1} \times \left(\frac{M}{0.01M}\right)^{1-\frac{\alpha}{2}} \left(\frac{v}{0.1c}\right)^{\frac{\alpha}{2}} \left(\frac{\kappa}{1\,\text{cm}^2\,\text{g}^{-1}}\right)^{-\frac{\alpha}{2}} \quad (3)$$

A larger ejecta mass produces a brighter and longer-lasting kilonova; a higher velocity gives a brighter and briefer kilonova.

The opacity of r-process ejecta—which arises from the blending of millions of bound-bound atomic line transitions—is complex in detail, but can be intuited from fundamental atomic physics[12]. For light r-process compositions, the most numerous lines are from iron group homologues with valence electrons in a *d*-shell orbital. Heavy r-process matter, however, contains a substantial fraction (1%–10% by mass) of lanthanides (atomic numbers $58 \leq Z \leq 71$) which have *f*-shell valence electrons. The complex *f*-shell species have more densely spaced energy levels and orders of magnitude more line transitions than the *d*-shell species. As a result, the opacity of heavy r-process ejecta is remarkably high, nearly 100 times that of typical astrophysical mixtures[12,29].

The models in Fig. 2 demonstrate the dramatic effect that composition has on kilonova emission. Ejecta composed primarily of light r-process material (lanthanide mass fractions ($X_{lan} \leq 10^{-4}$) has relatively low opacity ($\kappa \leq 1\,\text{cm}^2\,\text{g}^{-1}$) and radiates optical light that fades on a timescale of days. In contrast, ejecta composed primarily of heavy r-process elements ($X_{\text{lan}} \geq 10^{-2}$) has large opacity ($\kappa \approx 10\,\text{cm}^2\,\text{g}^{-1}$), a longer diffusion time (equation (2)) and light curves that last on the order of weeks. Lanthanide lines obscure the optical bands, and the radiation emerges primarily in the infrared.

Considering the observed properties of AT 2017gfo, we conclude that this merger produced two spatially distinct components of ejecta. The bright and brief optical light curve implies that one component of the ejecta was composed primarily of light (low-lanthanide) r-process ejecta. We can further deduce that this light r-process ejecta was expanding rapidly; Fig. 3 shows that models with modest speeds, $v_k = 0.03c$ (typical of ordinary supernovae), show conspicuous spectral line absorptions, while for higher velocities ($v_k$ = 0.1c–0.2c) the features broaden and blend into subtler undulations. That the optical spectra of AT 2017gfo were essentially featureless[20,21] implies velocities $v_k \approx 0.3c$, characteristic of dynamical merger ejecta.

The infrared light curve of AT 2017gfo indicates that the merger produced a second component of ejecta containing heavy r-process isotopes. The observed red colour and relatively long duration match the previously predicted[11,12] emission signatures of high-lanthanide ejecta (see Fig. 2), which are distinct from other known astrophysical transients[30]. Models lacking lanthanides do not have high enough opacity to explain emission lasting 2 weeks unless (see equation (2)) the ejected mass is unphysically large ($M \approx 1 M_\odot$, which would also predict too bright a peak luminosity) or the velocity was extremely low ($v \approx 10^{-3} c$, which would predict narrow spectral features not seen in observations).

The infrared spectroscopy of AT 2017gfo provides even more compelling evidence that the merger synthesized heavy r-process elements. Our models predict broad (approximately 0.2 µm) spectral bumps in the infrared (Fig. 3), a signature of emission lines of r-process material moving at approximately $0.1c$. Specifically, these models (and ones previously published[12]) predict spectral peaks appearing near 1.1 µm and 1.5 µm. Spectroscopy[17,31] of AT 2017gfo taken approximately 4 days after the merger found spectral peaks of similar width and wavelength position, in agreement with the theoretical expectations.

Although the features in the kilonova spectra are complex blends of many lines, we find that they encode information on the detailed chemical makeup. Figure 4 shows that the positions of the spectral peaks vary with the abundances of individual lanthanides. The main features in the model are primarily produced by neodymium ($Z = 60$), one of the most abundant lanthanides produced by the r-process. Whether the spectral peaks seen in AT 2017gfo can be robustly identified with neodymium is debatable given uncertainties in the current high-$Z$ atomic line data. Nevertheless, Fig. 4 demonstrates that, by improving the atomic inputs, a detailed chemical analysis of newly formed r-process material is in principle possible.

Summing the light from both the 'blue' (light r-process) and 'red' (heavy r-process) components of the ejecta provides a comprehensive theoretical model of AT 2017gfo (Fig. 5). The observed optical and infrared luminosities imply ejecta masses of $M_{blue} \approx 0.025 M_\odot$ and $M_{red} \approx 0.04 M_\odot$, respectively, while the spectral analysis indicates velocities of $v_{blue} \approx 0.3c$ and $v_{red} \approx 0.1c$. At early times ($t \lesssim 2$ d) the emission is then dominated by the 'blue' component and the spectra are largely featureless. At intermediate times ($t \approx 4$ d) the 'red' and 'blue' components emit at comparable strength, and the spectral energy distribution is a superposition of two quasi-blackbodies of different temperatures. Observations indeed show evidence for this distinctive 'double-peaked' spectrum at these epochs[15,16,20,21,31], confirming another specific theoretical prediction of a kilonova models[11]. At later times ($t \gtrsim 5$ d) the 'red' component dominates, and spectral wiggles are conspicuous in the infrared.

Our analysis paints a rich picture of GW170817, similar to that shown in Fig. 1b. The high velocity of the 'blue' kilonova component associates it with the dynamically 'squeezed' polar material[7,22,32]. The 'red' kilonova component could in part be due to tidally stripped material[4,32], but the inferred mass is rather large compared to simulations of dynamical ejecta, which rarely produce $M > 0.01 M_\odot$. Therefore it is likely that the 'red' component was produced in large part by a post-merger disk wind[6,26,33]. This suggests that the merged remnant may have collapsed quickly (less than about 10 ms) to a black hole, cutting off the neutrino irradiation that might otherwise lower the neutron fraction of the wind.

We anticipate that future kilonova observations will exhibit diversity. For binaries with a larger stellar mass ratio, the ratio of polar to tidal ejecta is predicted to be smaller, which would be measurable via the observed ratio of optical to infrared light. Neutron star–black hole mergers probably produce no polar ejecta, and so may be rather optically dim, but infrared bright (Fig. 1c). In systems that produce a lanthanide-free 'blue' disk wind (Fig.1a), we may see optical spectra with numerous spectral features[33] (see Fig. 3). Finally, orientation may play a part: if the geometry of Fig. 1b were to be viewed nearly edge-on, the optical emission from the poles could be partially obscured by the high-opacity tidal tails[33].

The substantial ejecta masses we infer from AT 2017gfo suggest that neutron-star mergers may be the dominant contributors to r-process production in the Galaxy. The discovery of an event in the LIGO O2 observing run is consistent with a relatively high rate of mergers in the Galaxy (see Methods). If the typical yields are indeed $M \geq 0.01 M_\odot$, the accumulated nucleosynthesis from mergers could account for all of the gold, platinum and many other heavy elements around us. The inferred ejecta mass of GW170817, however, should be considered approximate, owing to uncertainties in the radioactive heating rate, the uncertain atomic data used for opacities, and possible viewing angle effects (see Methods). Future theoretical and experimental work can address these limitations.

In the past, the uncertain origin of the heaviest elements was studied indirectly, by analysing fossil traces of these species in the surfaces of old stars. With AT 2017gfo we have now directly glimpsed and spectroscopically dissected a sample of pure r-process matter, big enough to enrich a million such stars. This astronomical phenomenon promises swift answers to old puzzles of cosmic origins.

**Acknowledgements** D.K. is supported in part by a Department of Energy Office (DOE) early career award DE-SC0008067, a DOE OfficeofNuclear Physics award DE-SC0017616, and by the Director, Office of Energy Research, Office of High Energy and Nuclear Physics, Divisions of Nuclear Physics, of the US

Department of Energy under contract number DE-AC02-05CH11231. This work was supported in part by the DOE SciDAC award DE-SC0018297. E.R.-R. acknowledges support from a Niels Bohr Professorship funded by DNRF, and support from UCMEXUS, the David and Lucile Packard Foundation. This research is funded in part by the Gordon and Betty Moore Foundation through grant GBMF5076. E.Q. was funded in part by the Simons Foundation through a Simons Investigator Award. J.B. is supported by the National Aeronautics and Space Administration (NASA) through the Einstein Fellowship Program, grant number PF7-180162, issued by the Chandra X-ray Observatory Center, which is operated by the Smithsonian Astrophysical Observatory for and on behalf of the National Aeronautics Space Administration under contract NAS8-03060. This research used resources of the National Energy Research Scientific Computing Center, a DOE Office of Science User Facility supported by the Office of Science of the US DOE under contract number DE AC02-05CH11231. .

**Author Contributions** D.K carried out the model calculations and analysis and led the writing of the manuscript. B.M. helped with the text, aided in the theoretical interpretation, and contributed to the schematic figure of mass ejection. J.B. carried out multi-dimensional radiation transport calculations to estimate the effects of asymmetry on the light curves. E.Q. provided theoretical interpretations and aided in the writing of the manuscript. E.R.-R. provided theoretical input and estimates of the contribution of mergers to the r-process in the Galaxy.

**Author Information** Reprints and permissions information is available at www.nature.com/reprints. The authors declare no competing financial interests. Readers are welcome to comment on the online version of the paper. Publisher's note: Springer Nature remains neutral with regard to jurisdictional claims in published maps and institutional affiliations. Correspondence and requests for materials should be addressed to D.K. (kasen@berkeley.edu).

**Reviewer Information** *Nature* thanks R. Chevalier and C. Miller for their contribution to the peer review of this work.

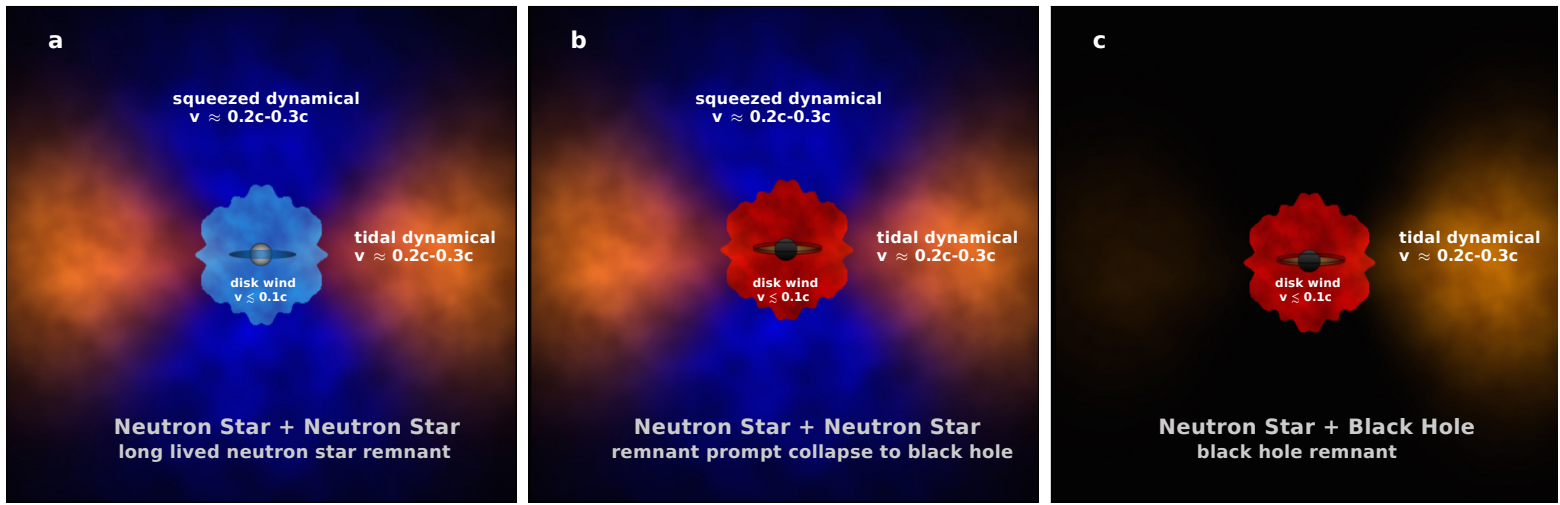

**Figure 1 | Schematic illustration of the components of matter ejected from neutron-star mergers.** Red colours denote regions of heavy r-process elements, which radiate red/infrared light. Blue colours denote regions of light r-process elements which radiate blue/optical light. During the merger, tidal forces peel off tails of matter, forming a torus of heavy r-process ejecta in the plane of the binary. Material squeezed into the polar regions during the stellar collision can form a cone of light r-process material. Roughly spherical winds from a remnant accretion disk can also contribute, and are sensitive to the fate of the central merger remnant. **a**, If the remnant survives as a hot neutron star for tens of milliseconds, its neutrino irradiation lowers the neutron fraction and produces a blue wind. **b**, If the remnant collapses promptly to a black hole, neutrino irradiation is suppressed and the winds may be red. **c**, In the merger of a neutron star and a black hole, only a single tidal tail is ejected and the disk winds are more likely to be red.

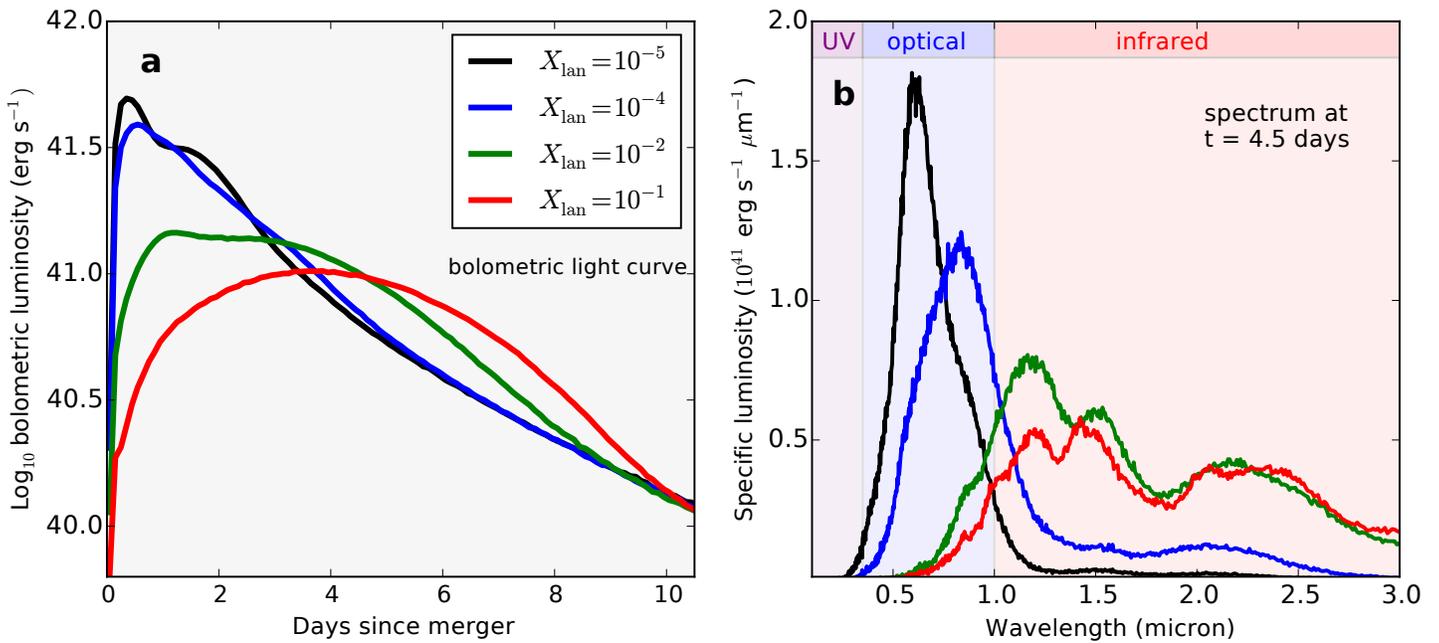

**Figure 2 | Models of kilonovae demonstrating the observable signatures of r-process abundances.** All models have an ejecta mass $M = 0.05 M_\odot$ and velocity $v_k = 0.2c$, but different mass fractions of lanthanides $X_{\text{lan}}$. **a**, Model bolometric light curves. If the ejecta is composed primarily of heavier r-process material ($X_{\text{lan}} \geq 10^{-2}$) the opacity is higher, resulting in a longer diffusion times and longer duration bolometric light curves. **b**, Model spectra as observed 4.5 d after the mergers. The higher lanthanide opacities of the heavy r-process materials obscure the optical bands and shift the emission primarily to the infrared.

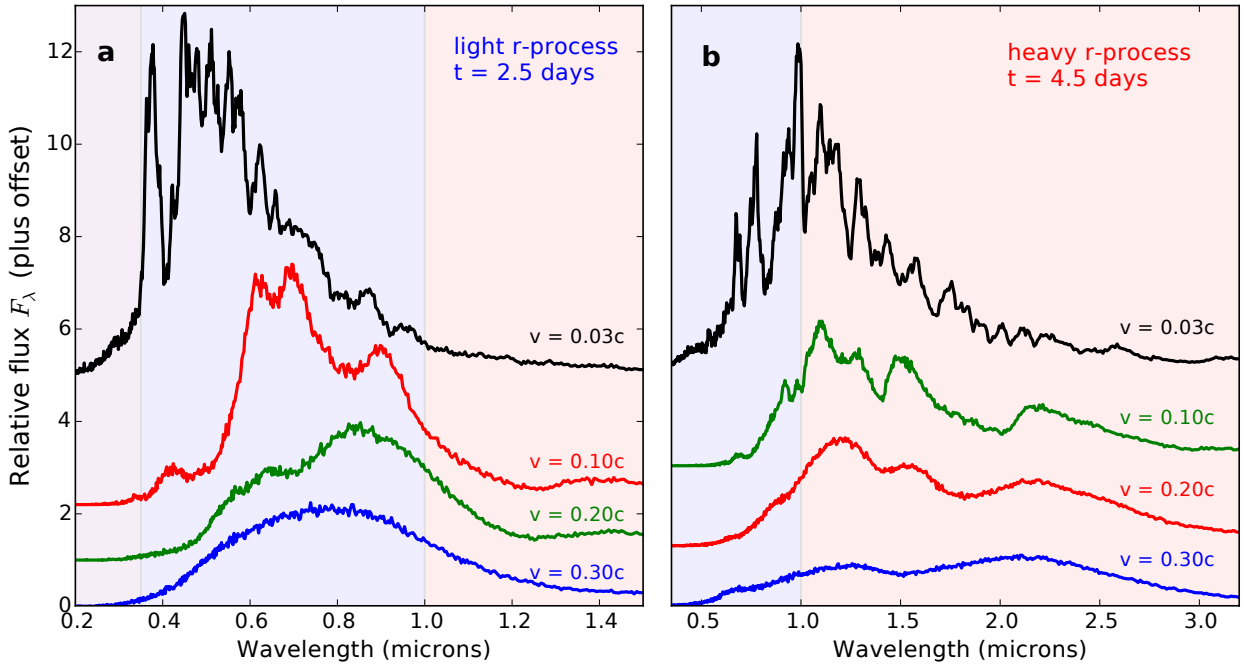

**Figure 3 | Models of kilonovae demonstrating the spectral diagnostics of the ejecta velocity.** The models all have ejecta mass $M = 0.03 M_\odot$. **a**, Spectra of models composed of light r-process material ($X_{lan} = 10^{-4}$) observed 1.5 d after the merger. Modest ejecta velocities ($v_k = 0.03c$, typical of supernovae) produce conspicuous absorption spectral features. At higher velocities ($v_k = 0.1c$–$0.2c$) the features are broadened and blended, while for $v_k = 0.3c$ the spectra are essentially featureless. The optical spectra of AT 2017gfo were featureless, implying a high-velocity, approximately $0.3c$ component of light r-process ejecta. **b**, Spectra of models composed of heavy r-process material ($X_{lan} = 10^{-2}$) observed 3.5 d after the merger. The infrared spectra of AT 2017gfo showed broad peaks, implying a lower-velocity, approximately $0.1c$ component of heavy r-process ejecta.

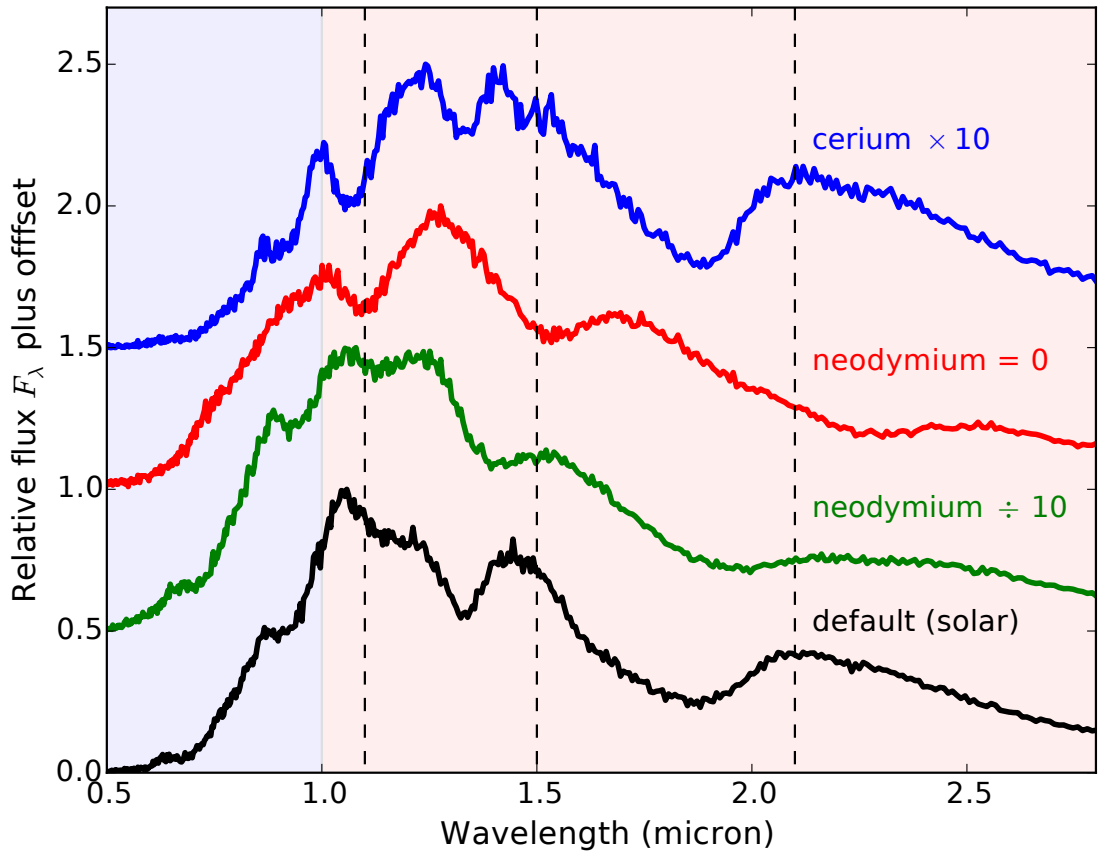

**Figure 4 | Models demonstrating how kilonova spectral features probe the abundance of individual r-process elements.** The spectral peaks in the models are blends of many lines, primarily those of the complex lanthanide species. The default model shown (parameters $M = 0.04 M_\odot$, $v_k = 0.15c$, $X_{lan} = 10^{-1.5}$) uses a solar distribution of lanthanides, and has spectral peaks near 1.1 μm, 1.5 μm and 2.0 μm (marked with dashed lines). These features are mainly attributable to neodymium ($Z = 60$) given that reducing or removing this species changes the feature locations. However, other lanthanides such as cerium ($Z = 58$) also affect the blended peaks. Uncertainties in the current atomic line data sources limit hinder spectral analysis, but with improved atomic inputs a more detailed compositional breakdown is within reach.

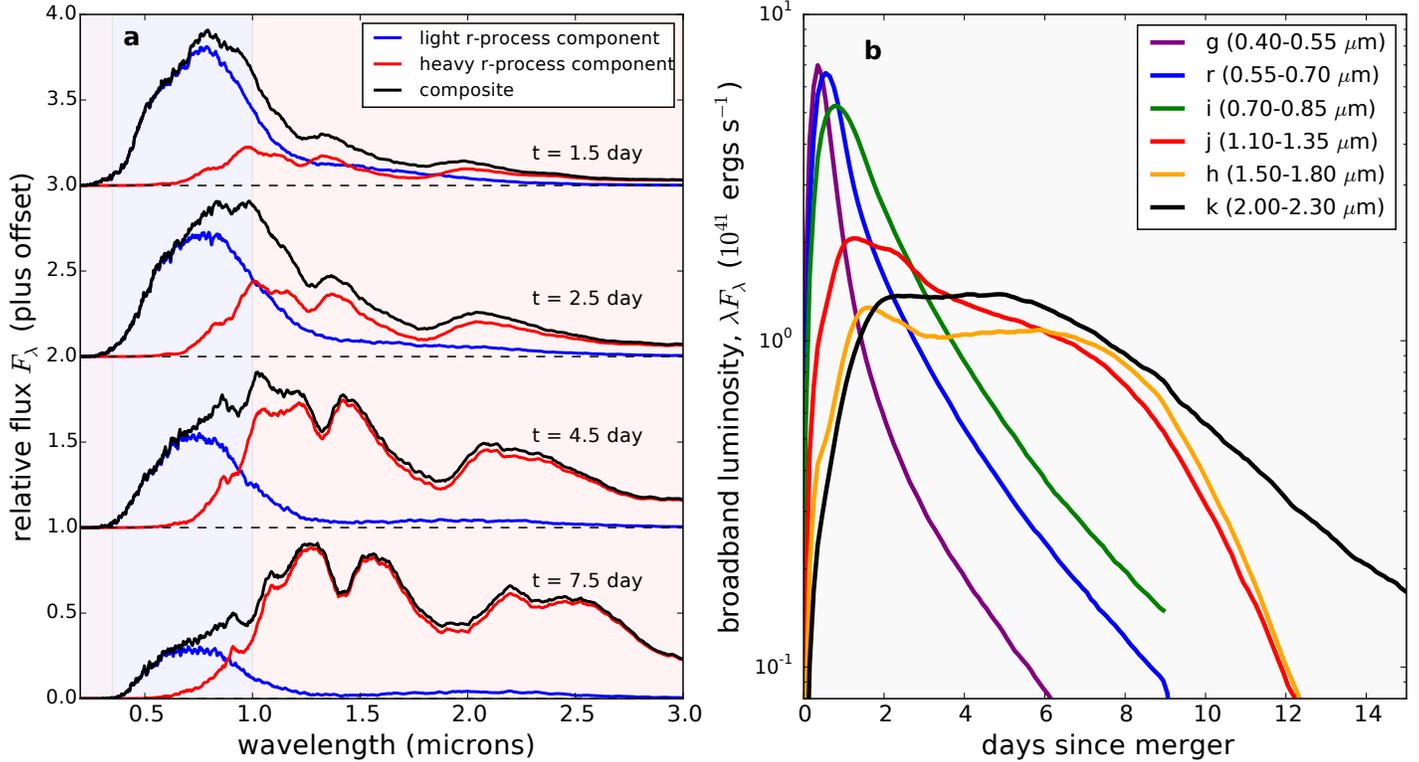

**Figure 5 | A unified kilonova model explaining the optical/infrared counterpart of GW170817.** The model is the superposition of the emission from two spatially distinct ejecta components: a 'blue' kilonova (light r-process ejecta with $M = 0.025 M_\odot$, $v_k = 0.3c$ and $X_{lan} = 10^{-4}$) plus a 'red' kilonova (heavy r-process ejecta with $M = 0.04 M_\odot$, $v_k = 0.15c$, and $X_{lan} = 10^{-1.5}$). **a**, Optical–infrared spectral time series, where the black line is the sum of the light r-process (blue line) and heavy r-process (red line) contributions. **b**, Composite broadband light curves. The light r-process component produces the rapidly evolving optical emission while the heavy r-process component produces the extended infrared emission. The composite model predicts a distinctive colour evolution, spectral continuum shape and infrared spectral peaks, all of which resemble the properties of AT 2017gfo.

## Methods

### Ejecta model parameterization

The kilonova light curves calculated here are based on parameterized ejecta models. The approach permits a controlled study of how kilonova observables depend on physical parameters, enabling a quantitative interpretation of observational data. The main parameters of the models are the ejecta mass $M$, the characteristic expansion velocity $v_k$, and the mass fraction of lanthanides $X_{lan}$, which largely controls the opacity. We surveyed model parameters in the range M = 0.001 – 0.1$M_\odot$ , $v_k$ = 0.03 – 0.4c, and $X_{lan}$ = 0 – 0.1, which are the ranges predicted by detailed merger simulations. For most models we assume spherical symmetry and a uniform composition, though for select models we explore the effects of non-spherical geometries and compositional gradients. Previous studies have calculated light curves based on realistic multi-dimensional dynamical models[33–36].

We initialize our calculations at a time $t_0$ = 0.1 d after the merger, well before the first observations of AT 2017gfo (at $t$ = 0.5 d) and early enough that radiative diffusion will not yet have substantially affected the ejecta. The ejecta can safely be assumed to be in free expansion at times $t_0 \gg t_{dyn}$ where the dynamical timescale at the time of ejection is $t_{dyn} \leq 1\ s$. The ejecta then evolve in a self-similar manner, with the radius proportional to the velocity, $r = vt$.

The ejected material is distributed over a range of radii and velocities, which we describe using a broken power-law density profile[37]. The density declines gradually in the interior but drops steeply in the outer layers

$$\rho(r,t) = \begin{cases} \zeta_\rho \dfrac{M}{v_t^3 t^3} \left(\dfrac{r}{v_t t}\right)^{-\delta} & \text{for } v < v_t \\ \zeta_\rho \dfrac{M}{v_t^3 t^3} \left(\dfrac{r}{v_t t}\right)^{-n} & \text{for } v \geq v_t \end{cases} \qquad (4)$$

where the transition velocity $v_t$ between the inner and outer layers is

$$v_t = \zeta_v \left(\dfrac{2E}{M}\right)^{1/2} \qquad (5)$$

with $E$ the kinetic energy of the ejecta. Our default values for the exponents are $\delta = 1$, $n = 10$, from which the normalization constants $\zeta_\rho$ and $\zeta_v$ can be determined by requiring that the density profile integrates to the specified mass and kinetic energy. We chose to write the energy $E$ in terms of a 'kinetic' velocity parameter

$$v_k = \left(\frac{2E}{M}\right)^{1/2} = \frac{v_t}{\zeta_v} \qquad (6)$$

that characterizes the velocity of the bulk of the ejecta. We take the outer edge of the simulation to be the point where the density has fallen to $10^{-3}$ of its value at the transition velocity $v_t$.

**Radiative transfer methods**

Having defined an ejecta model, we can calculate its light curves and spectra with no additional adjustable parameters. We do so using a multi-dimensional Monte Carlo code that solves the multi-wavelength radiation transport equation in a relativistically expanding medium[38,39]. In the Monte Carlo approach, the radiation field is represented by discrete particles that probabilistically sample the emission and absorption of photons as they propagate through the ejecta.

We use an analytical fit to the radioactive energy generation rate derived from detailed reaction network calculations of r-process isotopes[40]. This rate is assumed to be uniform throughout the ejecta and the same for all models. The fraction of radioactive decay energy $f_{th}(t)$ that is absorbed and thermalized in the ejecta is taken from the analytic fits derived from a detailed study[41] of kilonova thermalization efficiencies as a function of ejecta mass and velocity.

We calculate the temperature throughout the ejecta by balancing the local rate of radiative cooling with that of radiative and radioactive heating, and determine the ionization and excitation state of the plasma by solving the Saha–Boltzmann equations of local thermodynamic equilibrium. Though the densities in kilonova ejecta are too low to establish local thermodynamic equilibrium by collisional processes, the bulk of the ejecta is optically thick throughout much of its evolution. Repeated redistribution of photons via line transitions will drive the radiation field towards a blackbody distribution[42] and approximately establish local thermodynamic equilibrium level populations. At times well after than the light curve peak, when the ejecta becomes optically thin, the approximation of local thermodynamic equilibrium breaks down and our model spectra are less reliable.

To calculate the ejecta opacity, we use an atomic data set containing tens of millions of lines. The data for elements around the iron group ($Z = 21–28$) is extensive and well calibrated, but data for heavier species is sparse. Since homologous elements have very similar kilonova opacities[12] we use the line data for $Z = 21–28$ species as a stand in for the *d*-block r-process elements $Z = 39-48$ and $Z = 72–80$.

Previous kilonova transport calculations have included atomic data for a handful of lanthanide elements[11,33,36,43]. We use here new *ab initio* atomic structure models calculated with the Autostructure code[44] that include all of the complex lanthanides ($Z = 58–70$) with ionization states up to +4. While the *ab initio* structure models do not accurately predict the wavelength of individual lines, they do capture the statistical distribution of lines and so are adequate for estimating the kilonova opacity, which is the aggregate effect of great numbers of lines. For one element, neodymium, we have tuned the atomic structure models to roughly match the low-lying energy levels experimentally measured, such that the individual line wavelengths for this species are more reliable.

The large number of line transitions precludes an individual treatment. We therefore sum lines into frequency bins according to the Sobolev expansion opacity formalism[45,46]. The Sobolev approximation applies in the limit that the ion thermal velocities are small compared to the expansion velocities, which is well justified in merger ejecta. An alternative 'line-smeared' approach to binning lines has been proposed[47] that artificially broadens the linewidths by a large factor. This method fills in the low opacity windows between lines, which can substantially over-estimate the opacity. Radiation transport calculations[36] based on the line-smearing approach find somewhat redder light curves, which at maximum light cut off sharply at wavelengths less than 1.6 µm, unlike our models, which peak close to 1.1 µm.

In addition to atomic lines, we included opacity from electron-scattering and free-free processes, though the contribution of these sources is subdominant. Photons that interacted with lines were assumed to be absorbed and re-emitted according to a thermal emissivity, in keeping with the approximation of local thermodynamic equilibrium.

Our spherically symmetric models use 80 radial grid points, and 1,629 frequency bins logarithmically spaced between $\nu = 3 \times 10^{13}$ Hz and $\nu = 2 \times 10^{16}$ Hz. We begin the calculation at $t_0 = 0.1$ d, and end it at 30 d, and limit time steps to 10% of the elapsed time, which is sufficient to resolve the expansion evolution of the ejecta.

Monte Carlo photons that escape the ejecta are collected and binned in time and frequency to generate the spectral time series of the model, with all relevant Doppler shift and light travel time effects taken into account.

**Dependence on compositional gradients**

While our default ejecta models are chemically homogeneous, we expect realistic merger debris to show spatial variation in composition. In particular, since the neutrino luminosity following the merger typically declines over time, the interior layers of ejecta are probably more lanthanide-rich than the more highly irradiated exterior layers. Over time, as the matter expands and dilutes, observations probe progressively deeper layers of ejecta. If the lanthanide abundance increases radially inward, we anticipate a steady increase in line opacity.

To explore the effect of a compositional gradient, we consider a model of dynamical ejecta with $M = 0.025 M_\odot$ and $v_k = 0.25c$ and a lanthanide abundance

$$X_{\mathrm{lan}}(v) = (X_{\mathrm{lan,in}} - X_{\mathrm{lan,out}}) \left[ \left( \frac{v}{v_g} \right)^{n_g} + 1 \right]^{-1} + X_{\mathrm{lan,out}} \qquad (7)$$

where $X_{\mathrm{lan,in}}$ and $X_{\mathrm{lan,out}}$ are the asymptotic lanthanide mass fractions in the inner and outer layers, respectively, $v_g$ is the velocity coordinate around which the lanthanide abundance transitions between these values, and $n_g$ sets the sharpness of the transition.

Extended Data Fig. 1a shows optical light curves of a model with $X_{\mathrm{lan,in}} = 10^{-4}$, $X_{\mathrm{lan,out}} = 10^{-6}$, $v_g = 0.32c$ and $n_g = 12$. At early times, when the photosphere is in the outermost layers of ejecta, the light curves resemble those of a homogeneous model with $X_{\mathrm{lan}} = 10^{-6}$. A day later, the photosphere has receded into the layers of higher lanthanide abundance, and the light curve begins to more closely resemble a homogeneous model with $X_{\text{lan}} = 10^{-4}$.

The lanthanide gradient thus causes the model colours and optical light curves to evolve more rapidly than that of a homogenous model. The observations of AT 2017gfo show a rapid optical evolution[14] that can be better fitted with a lanthanide gradient[15,20,21]. This is an indication that, in principle, detailed modelling of the spectral evolution can probe the layered abundance stratification of a kilonova.

**Dependence on the ejecta density profile**

Our default models adopt a broken power-law density profile (equation (4)) with exponents $\delta = 1$ and $n = 10$. Overall, the model light curves are not particularly sensitive to the choice of power-law exponents or the detailed functional form of the profile. Using an exponential density profile, for example, gives quantitatively similar predictions.

Some specific observables do show substantial dependence on the density profile. The emission at the earliest epochs ($t \leq 1$ d) arises from the small amount of mass in the outer layers of ejecta. Models with a shallower outer profile ($n < 10$) have a larger, cooler photosphere and a redder spectrum. Extended Data Fig. 1b shows that choice of outer density profile strongly affects the ultraviolet luminosity at times $t \leq 1$ d. The relatively bright early ultraviolet emission of AT 2017gfo are more easily fitted[15] with a steep outer density drop-off ($n > 10$).

**Light curves of asymmetric models**

Simulations indicate that matter ejected in a neutron-star merger is probably aspherical, with light r-process material primarily distributed in the polar regions, and heavier tidal r-process ejecta concentrated in the equatorial plane (see Fig. 1). To explore the implications of asymmetry on the mass estimates for AT 2017gfo, we ran additional radiation transport calculations in two-dimensional cylindrical geometry. To describe the polar ejecta we used an *ad hoc* density profile

$$\rho_{\text{pol}}(r,\theta,t) = \rho(r,t)\left[1+\left(\frac{f(\theta)}{f_0}\right)^{10}\right]^{-1} \tag{8}$$

where $\rho(r, t)$ is the properly normalized spherical density profile (equation (4)) and $f(\theta) = 1 - \cos(\theta)$ and $f_0 = 1 - \cos(\theta_0)$. This formula concentrates the ejecta in the polar cone with half-opening angle $\theta_0$. We took the polar ejecta to have parameters $M = 0.025 M_\odot$, $v_k = 0.3c$, $X_{lan} = 10^{-5}$, $\theta_0 = 45°$. To describe the asymmetry of heavy r-process material, we adopted an oblate ellipsoid density distribution with aspect ratio $a = 4$ and model parameters: $M = 0.04 M_\odot$, $v_k = 0.1c$, $X_{lan} = 10^{-2}$.

Extended Data Fig. 2 shows the bolometric light curves of the aspherical models. We find that orientation effects are relatively minor for the polar ejecta alone, with the peak brightness varying by approximately 20% depending on the viewing angle. The equatorial ejecta show modest orientation effects by a factor of about 2. This oblate ellipsoid is brightest when viewed pole on, as the projected emitting surface area is greatest for this orientation.

We conclude that, for a single ejecta component, deviations from spherical symmetry likely only change the estimated masses by a factor of 2 or less. However, these calculations do not consider the more complicated interplay between the two ejecta components. Depending on the geometry and viewing angle, the radiation emitted by one ejecta component may be absorbed and reradiated at a different wavelength by another ejecta component. The resulting effects have been explored[33,36] in previous studies and can be further quantified in future targeted studies of AT 2017gfo.

**Neutron-star mergers as galactic r-process sites**

The neutron-rich ejecta from binary neutron-star mergers have long been proposed[1–3,48,49] as a solution for the uncertain sites of r-process nucleosynthesis[50,51]. Although several pieces of evidence have accumulated in recent years supporting this hypothesis[52–57], until now there has been little compelling and direct evidence for this association.

The substantial ejecta mass (at least a few hundredths of solar masses) inferred from AT 2017gfo suggest that the binary neutron-star mergers may be the dominant contributors to the production of heavy r-process elements in the Galaxy. The discovery of a binary neutron-star merger in the O2 observing run of LIGO–Virgo suggests a relatively high binary neutron-star volumetric rate of $R_{BNS} \geq 10^3$ Gpc$^{-3}$ yr$^{-1}$, consistent with the previous observing run O1 upper limits[58]. However, a more accurate estimate of the rate and its uncertainty will require a complete analysis by the LIGO/Virgo collaboration. Using the standard conversion[59] adopted by LIGO of 1 Milky Way Equivalent Galaxy (MWEG) per (4.4 Mpc)$^3$ this would imply a Milky Way merger rate of $N_{BNS} \geq 8 \times 10^{-5}$ yr$^{-1}$, slightly exceeding the range inferred by modelling the Galactic binary pulsar population[60,61].

To explain the total r-process abundances in the Milky Way stellar population, the required mass production rate of r-process nuclei is approximately[62] $M_{rp} \approx 10^{-6} M_\odot$ yr$^{-1}$. Of this total, roughly two-thirds is required to explain the light r-process ($A \leq 140$) with the remaining one-third required to explain the heavy r-process ($A \geq 140$). Although these required production rates are broadly consistent with observations[63,64], they come with several uncertainties, such as the total stellar content of the Milky Way and its halo as a function of metallicity (see ref. 65 for a discussion of some of the uncertainties).

For a Galactic binary neutron-star merger rate of about $8 \times 10^{-5}$ yr$^{-1}$, the ejecta mass per event required to explain the Galactic r-process production is about $10^{-2} M_\odot$ for the light-element r-process and $3 \times 10^{-3} M_\odot$ for the heavy-element component. The upper ranges of these estimates are consistent with the mass estimates we have derived from the red and blue components of the kilonova light curve of AT 2017gfo. However, if the true binary neutron-star rate is higher than the value we have adopted above, we would encounter the opposite problem of overproducing the Galactic r-process in stars, although a substantial fraction could remain in the gas phase.

Several uncertainties affect our inferences. These include possible errors in the estimated ejecta mass owing to uncertainties in the radioactive heating rate[41], uncertainties in the ejecta opacity owing to incomplete line-lists, and viewing angle effects[10,33,34,36]. In addition, our translation from Galactic r-process production rates to volumetric binary neutron-star rates depend on assumptions about what fraction of mergers occur near the Galactic plane as opposed to within the exterior halo, for which the r-process ejecta may be lost into the intergalactic medium[66,67].

Previous tentative evidence[52,53,68] of kilonova emission found in follow-up observations of γ-ray bursts implies similar masses (at least $0.01 M_\odot$) of r-process ejecta. The amount of heavy elements produced in AT 2017gfo may then not be atypical. However, future observations and analyses of multiple gravitational-wave sources and kilonovae will be required to pin down the binary neutron-star merger rate and mean heavy-element production.

**Code availability**

Codes used to generate model ejecta files and to synthesize the light curves plotted in the paper are available at https://github.com/dnkasen/Kasen_Kilonova_Models_2017 The code used to solve the radiative transport problem is not currently available as it is in the process of being readied and approved for public release.

**Data availability**

The datasets for all kilonova light curve and spectral time series models generated and displayed here are publicly available at https://github.com/dnkasen/Kasen_Kilonova_Models_2017.

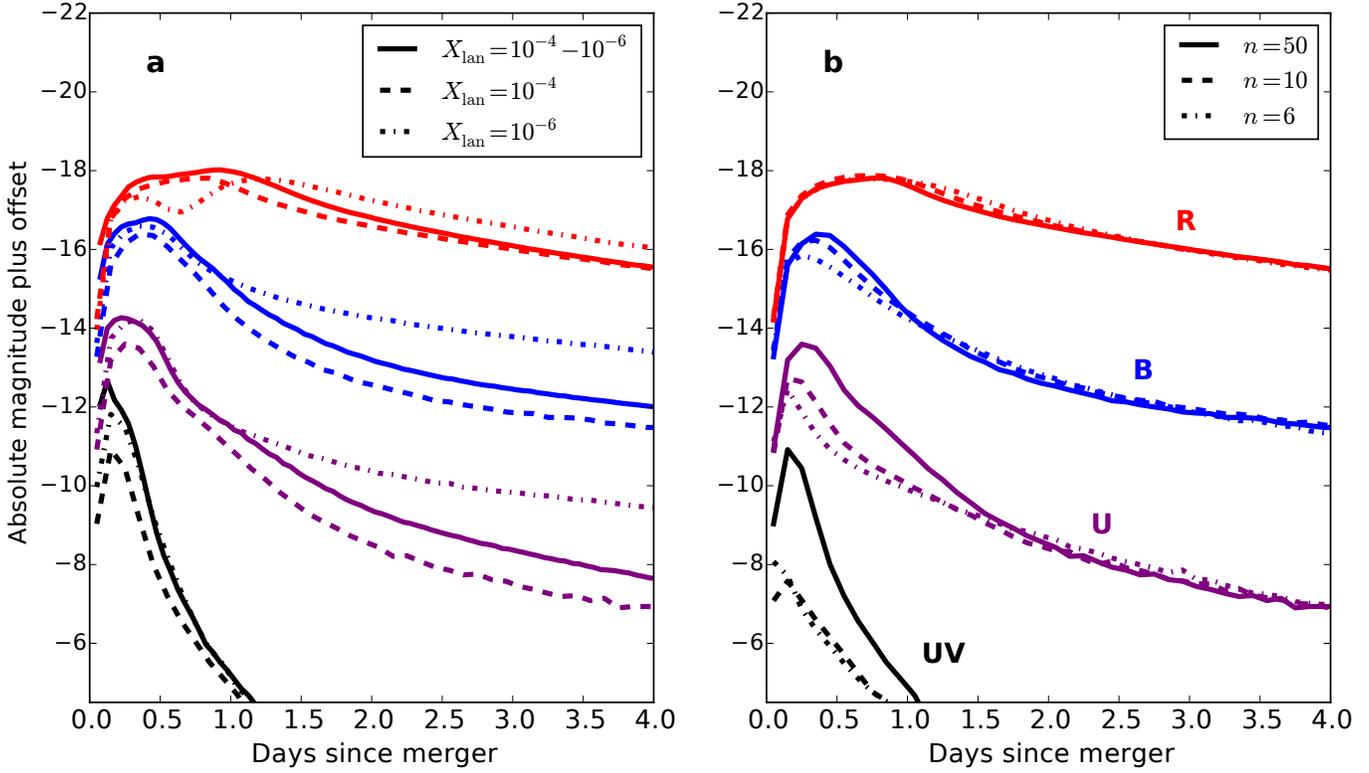

**Extended Data Figure 1 | Dependence of model light curves on the ejecta density profile and compositional stratification.** The models all have mass $M = 0.025 M_\odot$ and velocity $v_k = 0.25c$. **a**, Comparison of models with a homogenous composition to one where the lanthanide mass fraction varies from $X_{lan} = 10^{-6}$ at the outer ejecta edge to $X_{lan} = 10^{-4}$ in the interior (see equation (7)). **b**, Comparison of models with different density gradient in the outer layers. A shallower exponent ($n < 10$) leads to a cooler photosphere and suppresses the early ultraviolet and blue emission. The light curves at times $t \geq 1$ d and in redder bands are essentially independent of the outer density profile.

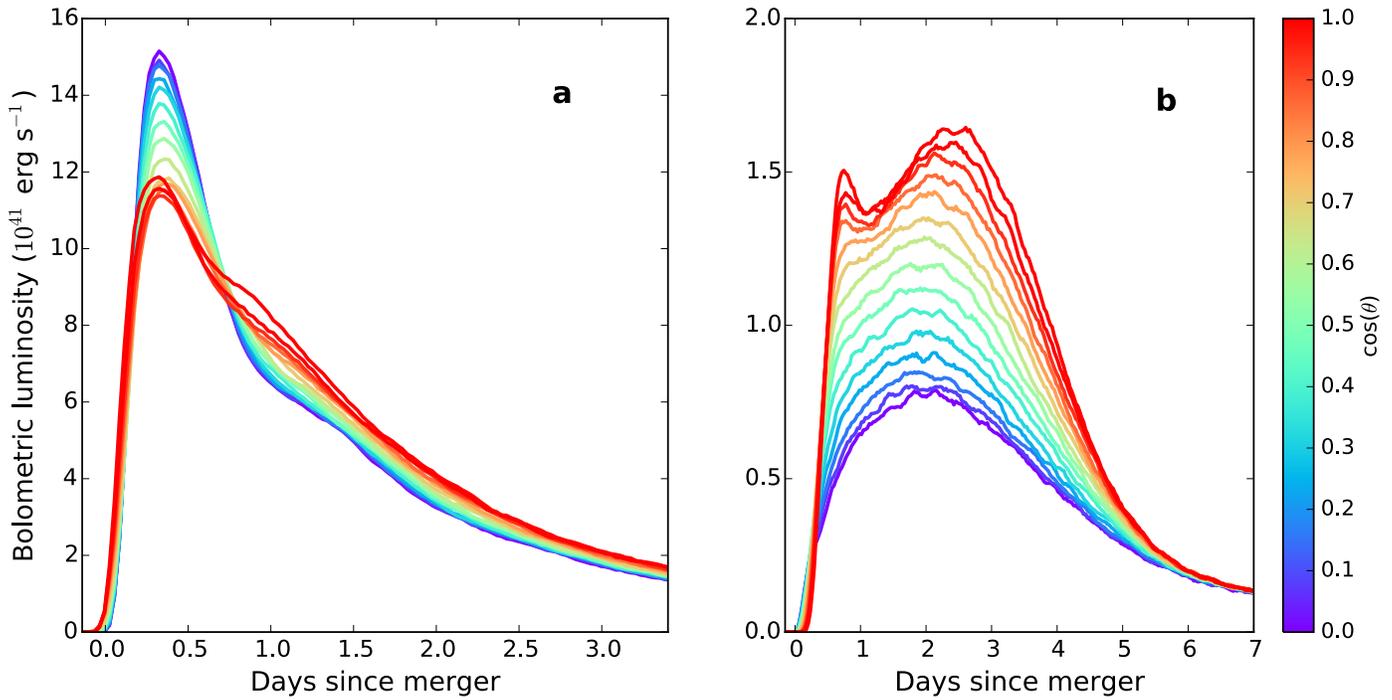

**Extended Data Figure 2 | Multi-dimensional models demonstrating the orientation dependence of asymmetric kilonova light curves. a**, Bolometric light curves of light r-process ejecta (with $M = 0.025 M_\odot$, $v_k = 0.15c$ and $X_{lan} = 10^{-5}$) distributed in a conical polar region of opening half angle 45°. **b**, Bolometric light curves of an oblate ellipsoidal distribution of heavy r-process ejecta (with $M = 0.04 M_\odot$, $v_k = 0.1c$ and $X_{lan} = 10^{-2}$) with an axis ratio of $a = 4$. The orientation effects lead to modest variations in the peak brightness. However, these models do not account for the effects of a superposition of the two multiple-component epochs.